\documentclass{aa}
\usepackage{graphicx}
\usepackage{xcolor}
\usepackage[fleqn]{amsmath}
\usepackage[utf8]{inputenc}
\usepackage{epsfig,amssymb}
\usepackage[varg]{txfonts}
\usepackage{multirow}
\usepackage{microtype}
\usepackage{subcaption}
\usepackage{placeins}
\usepackage{dblfloatfix}

\definecolor{myblue}{rgb}{0.00, 0.0, 0.9}
\definecolor{myred}{rgb}{0.90, 0.0, 0.0}
\definecolor{mygreen}{rgb}{0.0, 0.7, 0.0}
\usepackage[breaklinks,  citecolor=myblue, linkcolor=myred, urlcolor=purple, colorlinks=true, debug, baseurl=' ']{hyperref}
\usepackage{orcidlink}

\titlerunning{Zero-polarization candidate regions for the calibration of wide-field optical polarimeters}

\authorrunning{Mandarakas et al.}

\begin{document}

%

%
\title{Zero-polarization candidate regions for the calibration of wide-field optical polarimeters\thanks{Polarimetric data are only available in electronic form at the CDS via anonymous ftp to \url{cdsarc.u-strasbg.fr} (130.79.128.5) or via \url{http://cdsweb.u-strasbg.fr/cgi-bin/qcat?J/A+A/.}}}

\author{N.~Mandarakas
      \inst{1,2}\fnmsep\thanks{nmandarakas@physics.uoc.gr}\orcidlink{0000-0002-2567-2132},
      G.~V.~Panopoulou\inst{3}\orcidlink{0000-0001-7482-5759},
      V.~Pelgrims\inst{4}\orcidlink{0000-0002-5053-3847},
      S.~B.~Potter\inst{5,6}\orcidlink{0000-0002-5956-2249},
      V.~Pavlidou\inst{1,2},
      A.~Ramaprakash\inst{2,7,8},
      K.~Tassis\inst{1,2},
      D.~Blinov\inst{2},
      S.~Kiehlmann\inst{2}\orcidlink{0000-0001-6314-9177},
      E.~Koutsiona\inst{1},
      S.~Maharana\inst{5},\orcidlink{0000-0002-7072-3904}
      S.~Romanopoulos\inst{1,2},
      R.~Skalidis\inst{9}\orcidlink{0000-0003-2337-0277},
      A.~Vervelaki\inst{1}\orcidlink{0000-0003-0271-9724},
      S.~E.~Clark\inst{10,11}\orcidlink{0000-0002-7633-3376},
      J.~A.~Kypriotakis\inst{1,2}
      A.~C.~S.~Readhead\inst{2,9}
      }
          
\institute{Department of Physics, University of Crete, Vasilika Bouton, 70013 Heraklion, Greece
\and
Institute of Astrophysics, Foundation for Research and Technology – Hellas, 100 Nikolaou Plastira str. Vassilika Vouton, 70013
Heraklion, Crete, Greece
\and
Department of Space, Earth \& Environment, Chalmers University of Technology, SE-412 93 Gothenburg, Sweden
\and
Universit{\'e} Libre de Bruxelles, Science Faculty CP230, B-1050 Brussels, Belgium
\and
South African Astronomical Observatory, PO Box 9, Observatory, 7935, Cape Town, South Africa
\and
Department of Physics, University of Johannesburg, PO Box 524, Auckland Park 2006, South Africa
\and
Cahill Center for Astronomy and Astrophysics, California Institute of Technology, 1216 E California Blvd, Pasadena, CA, 91125, USA
\and
Inter-University Centre for Astronomy and Astrophysics, Post Bag 4, Ganeshkhind, Pune - 411 007, India
\and
Owens Valley Radio Observatory, California Institute of Technology, Pasadena, CA, 91125, USA
\and
Department of Physics, Stanford University, Stanford, CA 94305, USA
\and Kavli Institute for Particle Astrophysics \& Cosmology, P.O. Box 2450, Stanford University, Stanford, CA 94305, USA
}

\date{Received -- / Accepted --}

\abstract
   {The calibration of optical polarimeters relies on the use of stars with negligible polarization (i.e., unpolarized standard stars) for determining the instrumental polarization zero point. For wide-field polarimeters, calibration is often done by imaging the same star over multiple positions in the field of view (FoV), which is a time-consuming process. A more effective technique is to target fields containing multiple standard stars. While this method has been used for fields with highly polarized stars, there are no such sky regions with well measured unpolarized standard stars.}
   {We aim to identify sky regions with tens of stars exhibiting negligible polarization that are suitable for a zero-point calibration of wide-field polarimeters.}
   {We selected stars in regions with extremely low reddening, located at high Galactic latitudes. We targeted four $\sim 40' \times 40'$ fields in the northern and eight in the southern equatorial hemispheres. Observations were carried out at the Skinakas Observatory and the South African Astronomical Observatory.}
   {We found two fields in the north and seven in the south characterized by a mean polarization lower than $p<0.1\%$.}
   {At least 9 out of the 12 fields can be used for a zero-point calibration of wide-field polarimeters.}

\keywords{}

\maketitle


\section{Introduction}\label{sec:intro}

To date, the development of optical polarimetry instrumentation has primarily relied on the creation of instruments tailored for single-source observations or with limited field-of-view (FoV) capabilities. The calibration process for these instruments typically involves utilizing polarimetric standard stars \citep[e.g.,][]{Blinov2023}. When there is a need for calibrating the entire FoV, calibration methods often employ techniques such as rastering, where a star of known polarization is placed in different positions of the CCD to assess the instrument-induced polarization variations accross the field \citep[e.g.,][]{King2014}. This can be very time-consuming and, thus, inefficient for wide field instruments with a great deal of CCD pixels, especially in the scheme of wide-field surveys.

The domain of optical polarimetry is on the verge of a significant transition as it enters the era of large-area surveys such as \textsc{Pasiphae} \citep{Tassis2018}, SouthPol \citep{SouthPol}, and VSTPol \citep{Covino2020}. These surveys will utilize polarimeters equipped with unprecedentedly wide FoVs. However, the calibration process for these wide-field polarimeters presents a formidable challenge, particularly considering the surveys' aim of achieving high accuracy levels of $p \lesssim 0.1\%$ to fulfill their scientific objectives.

\textsc{Pasiphae} is a pioneering survey with the primary objective of mapping the polarization of millions of stars located away from the Galactic plane. The scientific goals of the project include the determination of dust cloud distributions and their properties along each line of sight (LoS), as well as the investigation of the magnetic field structure within our Galaxy \citep{Pelgrims2023}. Its findings will facilitate the accurate removal of the dust polarization foreground for surveys targeting the detection of the imprint of the primordial polarization B-modes on the cosmic microwave background. \citep[e.g.,][]{Polnarev1985,Kamionkowski1997}. \textsc{Pasiphae} will utilize two Wide-Area Linear Optical Polarimeters (WALOPs) which will operate from both the southern and northern hemispheres, mounted on the South African Astronomical Observatory's (SAAO) 1 m telescope\footnote{\url{https://www.saao.ac.za/}} and Skinakas Observatory 1.3 m telescope\footnote{\url{https://skinakas.physics.uoc.gr/}}, respectively. The description of the optical design, expected performance, and overall instrument design can be found in \cite{Maharana2020,Maharana2021}.

Given that the targeted accuracy of \textsc{Pasiphae} in polarization fraction is $0.1\%$, it is important to calibrate the whole FoV of WALOPs down to this level. \cite{Maharana2022} 
characterized the magnitude and sources of instrumental polarization and developed an on-sky polarimetric calibration method to obtain the target accuracy for WALOP-South. The calibration model creates an accurate mapping function between the instrument measured and the real Stokes parameters of a target. To develop the calibration model, wide-field sources of known and constant (also referred to as "flat") polarization across the field are observed through the instrument. The corresponding measured polarization values are used to create the mapping functions.
One such calibration source is the sky on bright-Moon nights in the vicinity of the Moon. \cite{Patat2006} calibrated the VLT-FORS1 instrument, using the qualitative assumption that the Moon-induced night sky polarization remains fairly constant in scales of a few arcminutes, while \cite{GonzalesGaitan2020} calibrated the FORS2 instrument by explicitly calculating the polarization of scattered moonlight.
\cite{Maharana2023} have demonstrated that the sky polarization remains constant at the level of $0.1\%$ or less for fields of size 10 to 20 arcminutes up to 20 degrees away from the Moon, making them suitable "polarimetric flat fields." As shown by \cite{Maharana2022}, the polarimetric flat fields suffice for calibration of wide field polarimeters, when used in conjugation with polarimetric standard stars. While polarimetric flat sources allow for relative calibration of the whole field, standard star measurements allow the estimation of the true polarization of the polarimetric flat source, enabling a complete and absolute calibration of the entire FoV.

Star fields with uniform polarization across the field can serve as candidate on-sky wide-field calibrators, provided their polarization scatter is smaller than the required calibration accuracy.
The advantages of this compared to the full Moon sky are: (1) it can be carried out on any given night, provided that such a field is available and (2) 
it can be used as standard calibrator source whose polarization is known directly from observations; this is unlike the full-Moon sky, whose polarization can only be calculated based on modeling or requires multiple observations at different angles from the moon \citep{GonzalesGaitan2020} to achieve a similar result to an observation of a zero-polarized field. This novel approach eliminates the need for additional standard star measurements needed to estimate the true polarization of the polarimetric flat source. A major (and often dominant) result of instrumental polarization that wide-field polarimeters tend to suffer from is the polarimetric zero offsets, namely: the measured Stokes parameters when the observed source is unpolarized. This needs to be corrected using unpolarized wide-field targets. 
By observing unpolarized targets, we can directly obtain the polarimetric zero offset, as any potential non-zero cross-talk terms between the Stokes $q=Q/I$ and $u=U/I$ are eliminated.
We note that in polarimeters that show non-negligible cross-talk between the linear and circular Stokes parameters, zero-polarization standards are not enough to address this issue. In such cases, the calibration would require more extensive modeling \citep{Wiersema2018}, observations of Stokes $V$ polarization standards \citep{Giro2003} and/or in-built calibration sources and/or devices within the instrument, as in the case of WALOP \citep{Maharana2022}.

Nonetheless, observations of zero-polarization fields are always essential.
To fill in this missing ingredient for calibrating wide-field polarimeters, here we aim to search for zero-polarized patches of the sky that can serve as calibrators for the WALOPs and other wide-field polarimeters. \cite{Skalidis2018} surveyed three $15' \times 15'$ regions of the northern sky at high Galactic latitudes with RoboPol and found one to be consistent with zero polarization, while \cite{Clemens2012} used polarized stellar cluster observations to calibrate the Mimir near-infrared imaging wide-field polarimeter \citep{Clemens2007}. In order to search for zero-polarized patches, we rely on the known connection between extinction by interstellar dust and observed polarization in a given LoS.

Interstellar dust is ubiquitous in every LoS \citep[e.g.,][]{Planck2014}. Its interaction with the interstellar magnetic field makes interstellar dust clouds act as polarizing filters. Asymmetric interstellar dust grains tend to align their minor axis parallel to the magnetic field lines \citep{Anddersson2015}. Dichroic absorption in optical wavelengths (i.e., the attenuation of the component of the light polarized in the plane-of-the-sky direction of the major axis of the grains) gives rise to dust-induced polarization to originally unpolarized sources behind said dust, with its orientation along the minor axis of the grains (i.e., along the magnetic field lines). In the infrared (IR) and sub-mm regime, dust emits light thermally, with the emission being polarized along the direction of its major axis, which is perpendicular to the polarization direction observed in the optical. The polarized emission from dust grains has been recently mapped with \textit{Planck} and shown to be well correlated with the optical starlight polarization \citep{Planck2020}.

In addition to dust-induced polarization, light passing through the dust suffers from reddening as well. The magnitude of these effects is related to the amount of dust along the LoS. Empirical investigations have established a  relationship between the maximum observed polarization $p_{max}$ and the reddening $E(B-V)$, expressed as $p_{max}=13\%E(B-V)$ \citep{Panopoulou2019,Planck2020}.
In this work we use this empirical relation, in conjunction with a publicly available reddening map to identify regions with expected polarization, $p\leq0.1 \%$, that could be used as calibrators for wide-field polarimeters.
We conducted polarimetric observations in the optical to verify whether these regions are indeed negligibly polarized. In Sect.~\ref{sec:data}, we describe the sample selection, data acquisition, and processing. 
In Sect.~\ref{sec:results}, we present the results of the polarization observations. We give our conclusions in Sect.~\ref{sec:disc} and provide supplementary plots in Appendices \ref{appendix:qu} and \ref{appendix:sky}.

\section{Data}\label{sec:data}

\subsection{Sample selection}\label{sec:sample}

Our goal is to target regions in the sky with minimal polarization and with an appreciable number of stars that can be used for determining the instrumental zero point across a FoV of $\sim40' \times 40'$, which is slightly wider than the FoV of the WALOPs that will be used for \textsc{Pasiphae}.
We are not only interested in finding specific zero-polarized stars in wide fields, such as fields with globular or open clusters, \citep[e.g.,][]{Clemens2012}. Such fields may indeed host clusters that are close enough so that their polarization is not affected by the interstellar medium (ISM). However, these fields may feature stars outside the cluster that are farther away and fully affected by the ISM. Thus, asserting that a certain region's polarization is minimally affected ensures that any given star in this region, regardless of distance, or brightness will be suitable for calibration, except in cases of intrinsically polarized stars, which can be easily identified. Thus, the calibration procedure becomes simpler, and the stars used for calibration do not necessarily need to be exactly the ones used in this specific work. For example, the use of other, potentially fainter, sources in our targeted regions would be equally qualified as calibrators, thus making these regions useful for bigger telescopes as well, where bright nearby targets would saturate their cameras.

Existing optical polarization data are sparse at high Galactic latitudes, with a mean density of 0.1 measurements per square degree \citep{Panopoulou2023}.
One avenue would be to identify regions with minimal polarized dust emission as observed by \textit{Planck}.  However, the \textit{Planck} polarization data are dominated by noise at high Galactic latitudes \citep{PlanckLFI2020,PlanckHFI2020}. Therefore, we searched for the lowest-polarization regions of the sky through reddening, which is an indirect observable of the expected polarization levels, as discussed in Sect.~\ref{sec:intro}. We used the \cite{Lenz2017} reddening map, as it is considered the most accurate at high Galactic latitudes \citep{Chiang2019}. Based on this map, we selected regions with $ E(B-V) < 0.01$ mag, corresponding to a maximum expected polarization fraction of  $p_{max} = 0.13\%$, according to the relation $p_{max}=13\%E(B-V)$ \citep{Panopoulou2019,Planck2020}. This resulted to 799 square degrees of usable sky area, compared to the total of 16,342 square degrees in the map.

Within the selected area, we searched for $40' \times 40'$ regions that fulfilled the maximum reddening of 0.01 mag criterion across the entire region. We aimed to find fields as spread-out with respect to the right ascension (RA) as possible, within the visibility constraints of the locations of the \textsc{Pasiphae} telescopes. This was to ensure that at least one such region is available on any observing night for calibration purposes. However, such regions are not available for as wide a range of RAs as we would wish for. Only four regions at northern latitudes and eight regions at southern latitudes available to our study satisfy the aforementioned criteria. Hereafter, we refer to these regions as "dark patches" (DPN for north, DPS for south).

Within each region, we selected all stars in \textit{Gaia} DR2 \citep{Gaia2016,gaia2018}, which was the latest available dataset at the time of target selection, with $G-$band magnitude $G < 14$ mag, to observe as many stars as possible in the least amount of time. Stars that were too bright ($G < 9$ mag) were excluded to avoid saturation on the CCD. We further excluded stars marked as variable in the \textit{Gaia} dataset. Stars with overlapping point spread functions, as seen in Digital Sky Survey (DSS)\footnote{\url{https://archive.eso.org/dss/dss}} images of the fields, were also excluded, as their polarization could be a result of spurious measurements.

\subsection{Data acquisition and reduction:\ North}
In order to acquire data for the dark patch candidates in the northern hemisphere, we utilized the RoboPol instrument, mounted on the Skinakas 1.3 m telescope in Crete, Greece. RoboPol
contains no rotating parts and can measure the linear Stokes parameters $q=Q/I$ and $u=U/I$ with a single exposure. Its novel design allows for  systematic and random errors to be minimized \citep{Ramaprakash2019}. The instrument is optimized for single target measurements in the center of its FoV, where the background noise is minimized with the use of a mask.

We observed four candidate patches between May and November 2021 and June to August 2023. The observations were conducted in Sloan Digital Sky Survey (SDSS)-r$'$ and Johnsons-Cousins R bands. Observations of both bands were considered together, as we verified with standard stars measurements that these filters are essentially equivalent \citep{Blinov2023}. The selection of these filters was also informed by the fact that the upcoming \textsc{Pasiphae} survey will be operating in SDSS-r$'$.
The data reduction and calibration were performed with the standard RoboPol pipeline, as outlined in \cite{King2014,Panopoulou2015,Blinov2021}. Uncertainties of the systematic error are on the order of $\sim0.1\%$ in the mask of the instrument \citep{Blinov2023}.
We note that these uncertainties concern measurements of individual targets. However, when we studied a particular field, the accuracy with of its mean polarization measurement is dependent on the number of stars observed in that field. The uncertainty of the mean polarization of the field decreases with the number of stars (refer to Eq.~\ref{eq:w.mean}) and may ultimately be lower than the uncertainty in the polarization of individual targets.

\subsection{Data acquisition and reduction:\ South}
All-Stokes photopolarimetry of the southern dark patches was performed during October of 2021 and August of 2022 on the 1.9-m telescope of the SAAO, using HIPPO \citep{2010MNRAS.402.1161P}. HIPPO's waveplates are contrarotated at 10 Hz and they therefore modulate the ordinary and extraordinary beams through the Thompson beamsplitter. The modulation is
sufficiently rapid that errors arising as a result of variable
atmospheric conditions or telescope guiding modulations become minimal. The systematic uncertainty of $\sim 0.1 \%$ was estimated from repeating and comparing observations of standards over multiple seasons. The modulated signal is sampled and recorded by two RCA31034A GaAs
photomultiplier tubes every millisecond. We refer to \cite{2010MNRAS.402.1161P} for more details. All measurements were performed with an R filter in the Kron–Cousin system. Polarized and unpolarized standard stars were observed to calculate the position angle offsets, instrumental polarization,
and efficiency factors. Background sky polarization measurements
were also taken at frequent intervals during the observations. The data
reduction then proceeded according to the method outlined in \cite{2010MNRAS.402.1161P}.

\section{Results}\label{sec:results}

All measurements of individual targets are corrected for the instrumental error as:
\begin{equation}
    q = q^{measured} - q^{inst},
\end{equation}
and similarly for $u$. The uncertainties of the individual targets include the contribution from instrumental error uncertainty. The final uncertainty is calculated as:
\begin{equation}\label{eq:w.mean}
    \sigma_q = \sqrt{ (\sigma_q^{measured})^2 + (\sigma_q^{inst})^2},
\end{equation}
and similarly for $u$, where the superscripts "measured" and "inst" refer to values obtained from differential photometry and the instrumental error, respectively. The instrumental error and its uncertainty are not necessarily identical for Stokes $q$ and $u$, nor between targets, as it can be time-dependent. The measurements of each targets were corrected using the values of the instrumental error corresponding to the same time period that the targets were observed.

\begin{figure*}[tbp]
    \centering
    \begin{subfigure}[b]{0.49\textwidth}
        \centering
        \includegraphics[width=\textwidth]{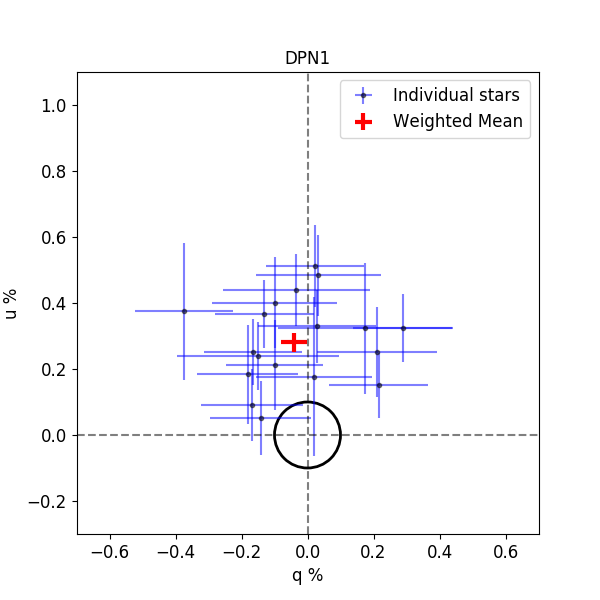}
        \label{fig:DPN1_qu}
    \end{subfigure}
    \hfill
    \begin{subfigure}[b]{0.49\textwidth}
        \centering
        \includegraphics[width=\textwidth]{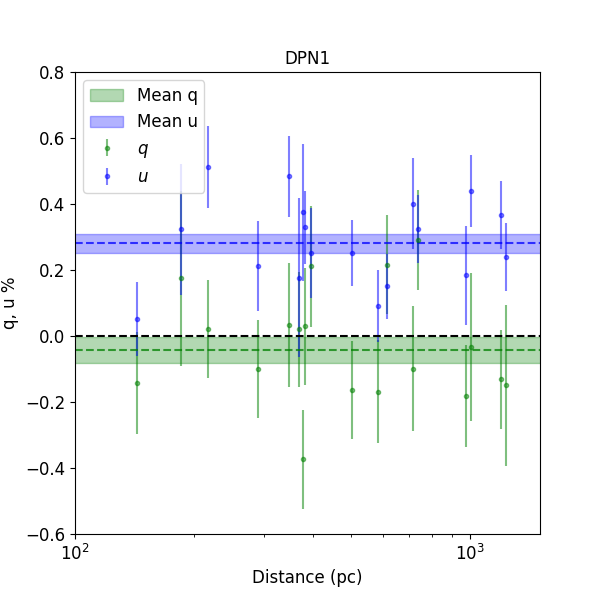}
        \label{fig:DPN1_dist}
    \end{subfigure}
    \caption{Polarization measurements of DPN1. Left: $q-u$ measurements of individual stars of DPN1 (blue), together with the weighted mean and the error on the weighted mean (red). The black circle marks the value $p=0.1\%$.  Right: $q$ (green) and $u$ (blue) measurements of the stars in the patch as a function of the distance from Earth. The green (blue) dashed line corresponds to the weighted mean $q$ ($u$) of the patch and the green (blue) shaded region to its error on the weighted mean.}
    \label{fig:DPN1}
\end{figure*}

\begin{figure*}[htbp]
    \centering
    \includegraphics[width=0.8\textwidth]{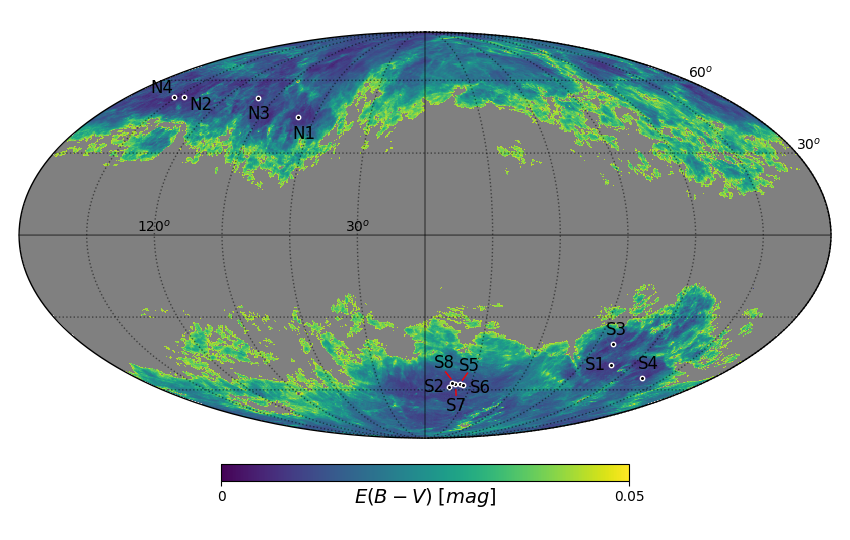}
    \caption{Locations of the regions surveyed in this work (white circles), overplotted on the \cite{Lenz2017} extinction map. The map is in Galactic coordinates, centered on l, b = (0,0) in Mollweide projection. Circle sizes are arbitrary. The red segments are used to distinguish between closely located fields. We use "N" instead of "DPN", and "S" instead of "DPS" for clarity. }
    \label{fig:wholeskymap}
\end{figure*}

We present the results of the dark patch candidates in Table~\ref{Tab:measurements}.
We present the $q-u$ measurements for DPN1 in Fig.~\ref{fig:DPN1}, together with their weighted mean, standard deviation, and error on the weighted mean. We also present $q,\,u$ as a function of distance in the same figure. Distance information was retrieved from the \textit{Gaia DR3} catalog \citep{Gaia2016,gaia2023}, by inverting the provided parallaxes. If parallax information was not available, we used the distance mentioned in the \textit{distance\_gspphot} column of the catalog. Stars with no distance information in either form were not plotted in the figure. Similar plots for all of the fields can be found in Appendix~\ref{appendix:qu}. The sky location of the dark patches are presented in Fig.~\ref{fig:wholeskymap}. Finder charts of the individual stars for which we measured polarization in each region are presented in Appendix~\ref{appendix:sky}. For each field, we divided the range of RAs and Decs. of our targets in four equal bins. For each bin, we calculated and plotted the average polarization $\bar{p}_{bin}$, and its standard deviation. For all of the regions, there is no significant deviation in the polarization in any of the bins. Therefore, there is no spatial dependence of the polarization within the fields and it can be considered constant in terms of polarization.
The sky maps were obtained from DSS.

Individual stellar measurements of $q,\, u$ have average statistical uncertainties of $\bar{\sigma q} = 0.175\%$ and $\bar{\sigma u} = 0.133\%$ for the northern fields, and $\bar{\sigma q}=0.097\%$ and $\bar{\sigma u} = 0.097\%$ for the southern.
However, the values that characterize the average polarization of each field is the weighted mean of the measurements and the error on the weighted mean, $\bar{q} \pm \sigma_{\bar{q}}$ and $\bar{u} \pm \sigma_{\bar{u}}$.
Therefore, for each field, we calculate the weighted mean of the Stokes parameters as: 
\begin{equation}\label{eq:w.mean}
    \bar{q} = \frac{\sum_{i=1}^{N} w_i \cdot q_i}{\sum_{i=1}^{N} w_i},
\end{equation}
with $i$ being the individual measurement of each star and $N$ the total number of stars in each field, with $w_i=1/(\sigma_{qi})^2$, and $\sigma_{qi}$ the uncertainty of the measurement. We are also interested in the weighted standard deviation, defined as:
\begin{equation}\label{eq:w.mean}
    q_{std} = \sqrt{\frac{\sum_{i=1}^{N} w_i \cdot (q_i - \bar{q})^2}{\sum_{i=1}^{N} w_i}},
\end{equation}
and the error of the weighted mean, which, in the case of measurements with different uncertainties (as ours) is: 

\begin{equation}\label{eq:w.mean}
    \sigma_{\bar{q}} = \frac{1}{\sqrt{\sum_{i=1}^{N} w_i}},
\end{equation}
and similarly for $u$.

Polarization is a positively defined quantity, expressed as:
\begin{equation}\label{eq:p,sp}
    p=\sqrt{q^2+u^2} \text{, with its uncertainty } \sigma_p=\sqrt{\frac{q^2 \sigma_q^2 + u^2 \sigma_u^2}{q^2+u^2}},
\end{equation}
with $\sigma_q$ and $\sigma_u$ being the uncertainties of $q$ and $u,$ respectively.
Therefore, measurements are biased towards higher values, especially for low signal-to-noise (S/N) measurements \citep[e.g.,][]{Vaillancourt2006}.
We debiased our measurements using the modified asymptotic (MAS) estimator proposed in \cite{Plaszczynski2014}: 
\begin{equation}
    p_{MAS} = p - \sigma_p^2 \frac{1-e^{-p^2/\sigma_p^2}}{2p}
,\end{equation}
with $p$ and $\sigma_p$ defined in Eq.~\ref{eq:p,sp}. We present the debiased values in Table~\ref{Tab:measurements}. 
Conversely, the bias in the polarization angle is likely very small \citep{Montier2015}. Thus, the expression of the polarization angle $\chi=\frac{1}{2} \text{arctan}(\frac{u}{q})$ was used without correction.

The observed polarization of stars in any given field is expected to arise solely due to the dichroic absorption of the light by aligned dust grains (as discussed in Sect.~\ref{sec:intro}). This is based on the fact that most stars are intrinsically unpolarized with the possible exceptions of magnetic stars or stars surrounded by dusty disks \citep{Fadeyev2007,Clarke2010}. The potential existence of such stars in our samples would manifest as clear outliers in the $q-u$ plots. For each of our patches, we performed a $3-\sigma$ clipping three times to discard outliers. This process excluded only 1 star from DPN3, 1 from DPN4, and 1 star from DPS7, while the number of the remaining stars per patch ranges between 16 and 31.

The observed scatter between the measurements of different stars in the same region is a combination of the intrinsic scatter of the sources within the region and the statistical scatter introduced by the instrument, arising from photon noise and instrumental systematics, namely: 
\begin{equation}
    \sigma_{q}^{observed} = \sqrt{\left(\sigma^{intrinsic}_{q}\right)^2 + \left(\sigma_{q}^{statistical}\right)^2} ,
\end{equation}
and it is similar for $u$, with the assumption that these sources of variance are Gaussian. The statistical scatter of both the northern and the southern instruments used for this study is on the order of $\sim0.1\%$, as deduced by standards observations. Therefore, the intrinsic scatter in the polarization of the observed patches is maximum $0.1\%$ for all cases.

\begin{table*}[tbp]
\caption{Properties of the dark patch candidates.}
\centering
\begin{tabular}{c c c c c c c c c c c c c c}
\hline
\hline
Name & $N_{stars}$ & RA & Dec & $\bar{p}$ & $\sigma_{\bar{p}}$ & $\bar{\chi}$ & $\sigma_{\bar{\chi}}$ & $\bar{q}$ & $q_{std}$ & $\sigma_{\bar{q}}$ & $\bar{u}$ & $u_{std}$ & $\sigma_{\bar{u}}$ \\
 & & (deg) & (deg) & ($\%$)  & ($\%$) & ($^{\circ}$) & ($^{\circ}$) & ($\%$) & ($\%$) & ($\%$) & ($\%$) & ($\%$) & ($\%$)
\\
\hline
\multicolumn{13}{c}{North} \\
\hline
DPN1 & 17 & 246.7065 & 44.0438 & 0.283 & 0.029 & 49.236 & 3.268 & -0.042 & 0.179 & 0.039 & 0.282 & 0.134 & 0.029 \\

DPN2 & 25 & 165.3404 & 59.5480 & 0.105 & 0.026 & 31.952 & 8.513 & 0.048 & 0.202 & 0.032 & 0.098 & 0.140 & 0.025 \\

DPN3 & 16 & 220.4940 & 59.9686 & 0.058 & 0.041 & 82.982 & 22.315 & -0.068 & 0.177 & 0.041 & 0.017 & 0.102 & 0.032 \\

DPN4 & 31 & 160.4000 & 56.6931 & 0.049 & 0.023 & 50.342 & 14.019 & -0.010 & 0.099 & 0.027 & 0.053 & 0.108 & 0.023 \\
\hline 
\multicolumn{13}{c}{South} \\
\hline
DPS1 & 29 & 59.1394 & -45.6599 & 0.093 & 0.018 & 36.449 & 6.675 & 0.028 & 0.098 & 0.018 & 0.091 & 0.119 & 0.018 \\

DPS2 & 37 & 342.303 & -47.6786 & 0.123 & 0.014 & -43.864 & 6.829 & 0.005 & 0.086 & 0.014 & -0.126 & 0.134 & 0.014 \\

DPS3 & 25 & 70.4262 & -53.1910 & 0.039 & 0.019 & 26.565 & 18.178 & 0.027 & 0.093 & 0.019 & 0.036 & 0.145 & 0.019 \\

DPS4 & 23 & 53.0764 & -28.4748 & 0.047 & 0.020 & 42.145 & 12.009 & 0.005 & 0.110 & 0.020 & 0.050 & 0.115 & 0.020 \\

DPS5 & 22 & 343.9268 & -51.2097 & 0.049 & 0.020 & -70.166 & 12.772 & -0.041 & 0.098 & 0.020 & -0.034 & 0.090 & 0.020 \\

DPS6 & 26 & 345.6808 & -51.7997 & 0.049 & 0.018 & -63.652 & 12.131 & -0.032 & 0.114 & 0.018 & -0.042 & 0.102 & 0.018 \\

DPS7 & 25 & 342.3994 & -49.7344 & 0.024 & 0.020 & -63.435 & 24.630 & -0.018 & 0.117 & 0.020 & -0.024 & 0.090 & 0.020 \\

DPS8 & 24 & 340.8273 & -49.0403 & 0.069 & 0.019 & -65.817 & 9.862 & -0.048 & 0.106 & 0.019 & -0.054 & 0.119 & 0.019 \\
\hline 
\multicolumn{13}{c}{Dark patches of \cite{Skalidis2018}} \\
\hline
DP1 & 24 & 191.2318 & 57.1060 & 0.0 & 0.038 & 41 & 22 & 0.007 &  & 0.041 & 0.053 &  & 0.037 \\

DP2 & 23 & 150.7294 & 54.4580 & 0.107 & 0.036 & 18 & 9 & 0.091 &  &0.036  & 0.066 &  & 0.036 \\

DP3 & 21 & 144.9987 & 34.1192 & 0.203 & 0.044 & 6 & 5 & 0.203 &  & 0.045 & 0.045 &  & 0.037 \\
\hline
\end{tabular}
\tablefoot{$\bar{p}$ and $\sigma_{\bar{p}}$ correspond to the debiased estimate of the mean polarization, and its error, respectively; $\bar{\chi}$ and $\sigma_{\bar{\chi}}$ are similar for the polarization angle; $\bar{q}$, $q_{std}$, and $\sigma_{\bar{q}}$ are the weighted mean of $q$ values, its standard deviation, and error on the mean respectively, and a similar meaning for $u$. Empty cells correspond to information that is not available.}
\label{Tab:measurements}
\end{table*}

\section{Discussion}\label{sec:disc}

We have identified and surveyed twelve regions of the sky with very low extinction values to assess whether they exhibit a sufficiently low polarization to be used as calibrators for wide-field polarimeters. We measured the linear polarization of stars in four patches in the northern hemisphere and eight in the southern. We find that two regions in the north and seven in the south display mean polarization of $\bar{p}<0.1\%$, while one more in each hemisphere has a mean polarization that is marginally above $0.1\%$. Moreover, we have included the regions probed by \cite{Skalidis2018} in Table~\ref{Tab:measurements} as one of those regions could also serve as a zero-polarization calibrator.

At high Galactic latitudes (i.e., the regions in our study), the full column density, as well as the dust polarization in emission and absorption are dominated by the wall of the Local Bubble located between 150 and 300 $pc$ from the Sun \citep{Skalidis2019,Pelgrims2020}. Most of the observed stars in our samples have distances larger than that.
Therefore, we are confident that our stellar polarization measurements sample all the dust column that could induce polarization along the LoS of the regions surveyed (see plots in Appendix~\ref{appendix:qu}). 
Moreover, we do not notice any significant change of the measured polarization as a function of distance. In conclusion, our measurements are characteristic of each field, regardless of distance, and represent the true polarization value arising due to the effect of interstellar dust in each region.

An interesting outcome of this study is the mean polarization of DPN1.
Based on its reddening value from \cite{Lenz2017} $E(B-V) \sim 0.06$ mag and the relation $p_{max}=13\%E(B-V)$, we would not expect its polarization to be at the measured value $p=0.283\pm 0.029\%$. This could be due to reasons related to how the dust maps are constructed and their uncertainties. There is quite high uncertainty in the modeling that is associated with dust maps at high latitude. For example, \cite{Lenz2017} compared the \cite{SFD1998} maps with $H_1$ data \citep{HI4PI2016} to derive the reddening $E(B-V)$.
This technique of modeling involves uncertainties relating to two factors: i) the gas-to-dust ratio, which may exhibit variations across the sky \citep{Shull2023,Skalidis2023}, and  ii) the dust modeling, which results in important differences at high latitudes \citep{Chiang2019,Chiang2023}. 
Characteristic evidence for the high uncertainties in the dust maps is revealed by the differences between maps, especially at high latitudes.
For instance, while the reddening value for DPN1 from \cite{Lenz2017} is $E(B-V) \sim 0.006$ mag, the corresponding value derived from the \cite{Planck2016} map is $E(B-V) \sim 0.013$ mag.
At the same time, the stellar reddening based maps \citep[e.g.,][]{Green2019} have statistical uncertainties that are comparable to the extinction, so they too cannot be trusted on a pixel-by-pixel basis.
Having discussed this peculiarity, it is worth mentioning that the recent work of \cite{Angarita2023} placed higher constraints on the relation $p_{max}/E(B-V)$, finding an upper limit of $p_{max}\sim 16\%E(B-V)$ in intermediate latititudes $|b|>7.5 \deg$.
Therefore, the extraordinary value in the mean polarization of DPN1 seems to be mostly associated with the uncertainty of the extinction of the field, but it is not impossible for it to be an outlier in the $p_{max}/E(B-V)$ relation.

In conclusion, the most important aspect of this work is that all fields, even the ones with $\bar{p}>0.1\%$, display very low intrinsic scatter in polarization ($<0.1\%$). Thus, they can be considered as polarimetric standard fields.
Depending on the needs and aims in accuracy of different polarization surveys and instruments, we expect that a total between 9 and all 12 of the surveyed regions could be used as zero-polarization calibrators for wide-field polarimeters.
All the data in this study are to be made publicly available on CDS.

\begin{acknowledgements}
The authors thank A. Steiakaki for useful discussions. We acknowledge the assistance of the anonymous referee who helped make this manuscript better.
The \textsc{Pasiphae} program is supported by grants from the European Research Council (ERC) under grant agreements No. 771282 and No. 772253; by the National Science Foundation (NSF) award AST-2109127; by the National Research Foundation of South Africa under the National Equipment Programme; by the Stavros Niarchos Foundation under grant \textsc{Pasiphae}; and by the Infosys Foundation.
VPa and SR acknowledge support by the Hellenic Foundation for
Research and Innovation under the “First Call for H.F.R.I. Research Projects to support Faculty members and Researchers and the procurement of high-cost research equipment grant”, Project 1552 CIRCE, and by the Foundation
of Research and Technology – Hellas Synergy Grants Program (project MagMASim)
VPe acknowledges funding from a Marie Curie Action of the European Union (grant agreement No. 101107047).
This work was supported by NSF grant AST-2109127.
%
%

 
The Digitized Sky Surveys were produced at the Space Telescope Science Institute under U.S. Government grant NAG W-2166. The images of these surveys are based on photographic data obtained using the Oschin Schmidt Telescope on Palomar Mountain and the UK Schmidt Telescope. The plates were processed into the present compressed digital form with the permission of these institutions. 

 This work has made use of data from the European Space Agency (ESA) mission
{\it Gaia} (\url{https://www.cosmos.esa.int/gaia}), processed by the {\it Gaia}
Data Processing and Analysis Consortium (DPAC,
\url{https://www.cosmos.esa.int/web/gaia/dpac/consortium}). Funding for the DPAC
has been provided by national institutions, in particular the institutions
participating in the {\it Gaia} Multilateral Agreement.

\end{acknowledgements}

\bibliographystyle{aa}
\bibliography{bibliography}

\appendix

\section{Polarization measurements of the observed fields}\label{appendix:qu}
We present our measurements of all of the dark patches, similarly to Fig.~\ref{fig:DPN1}.

\begin{figure*}[tbp]
    \centering
    \begin{subfigure}[b]{0.49\textwidth}
        \centering
        \includegraphics[width=\textwidth]{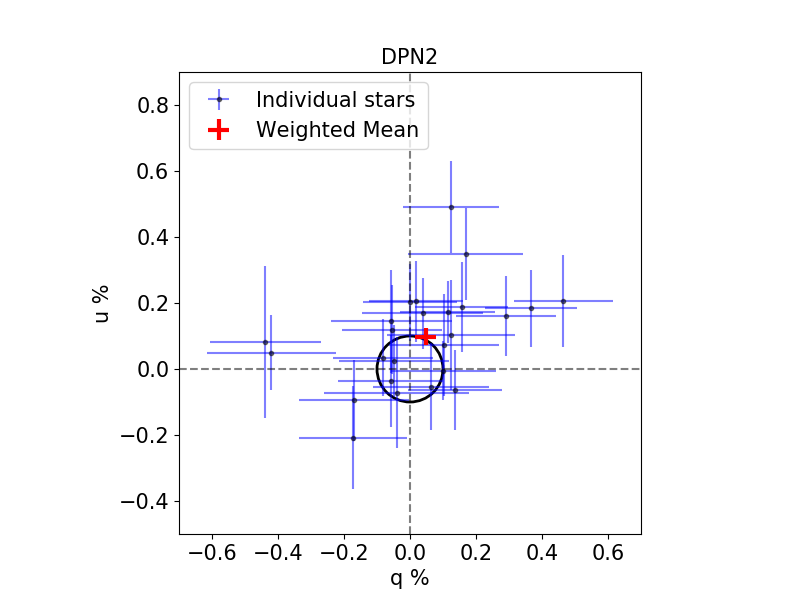}
        \label{fig:DPN2_qu}
    \end{subfigure}
    \hfill
    \begin{subfigure}[b]{0.49\textwidth}
        \centering
        \includegraphics[width=\textwidth]{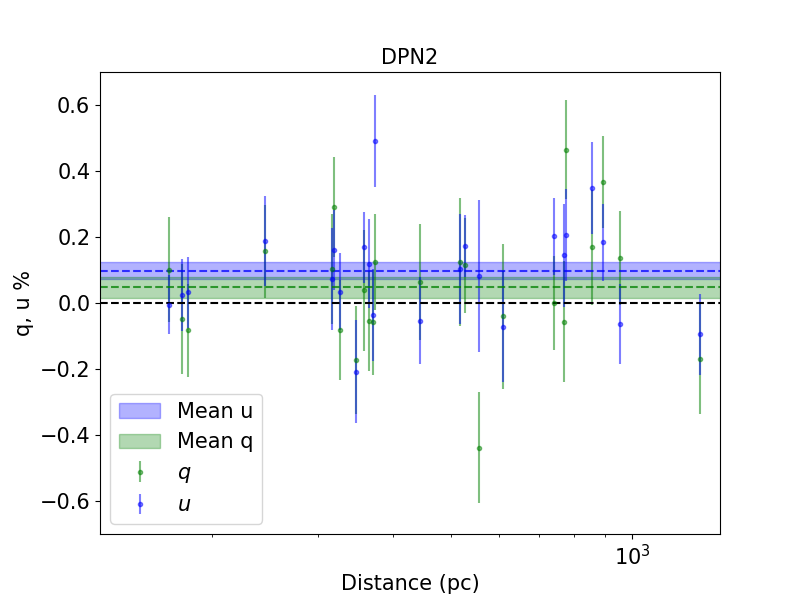}
        \label{fig:DPN2_dist}
    \end{subfigure}
    \caption{Same as Fig.~\ref{fig:DPN1} but for DPN2.}
    \label{fig:DPN2}
\end{figure*}

\begin{figure*}[tbp]
    \centering
    \begin{subfigure}[b]{0.49\textwidth}
        \centering
        \includegraphics[width=\textwidth]{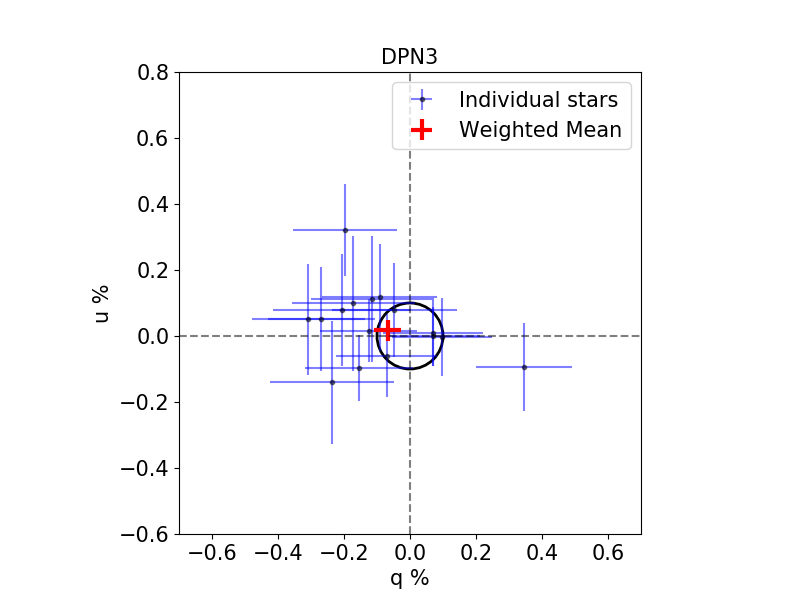}
        \label{fig:DPN3_qu}
    \end{subfigure}
    \hfill
    \begin{subfigure}[b]{0.49\textwidth}
        \centering
        \includegraphics[width=\textwidth]{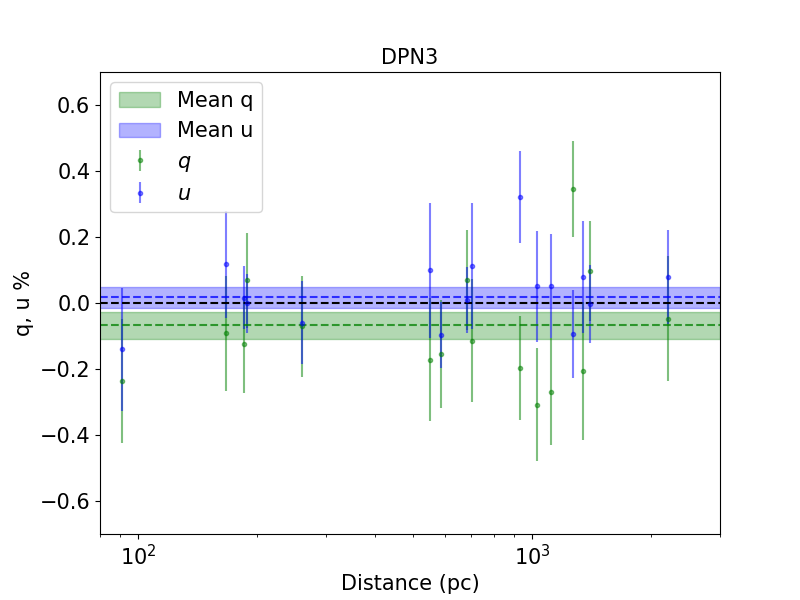}
        \label{fig:DPN3_dist}
    \end{subfigure}
    \caption{Same as Fig.~\ref{fig:DPN1} but for DPN3.}
    \label{fig:DPN3}
\end{figure*}

\begin{figure*}[tbp]
    \centering
    \begin{subfigure}[b]{0.49\textwidth}
        \centering
        \includegraphics[width=\textwidth]{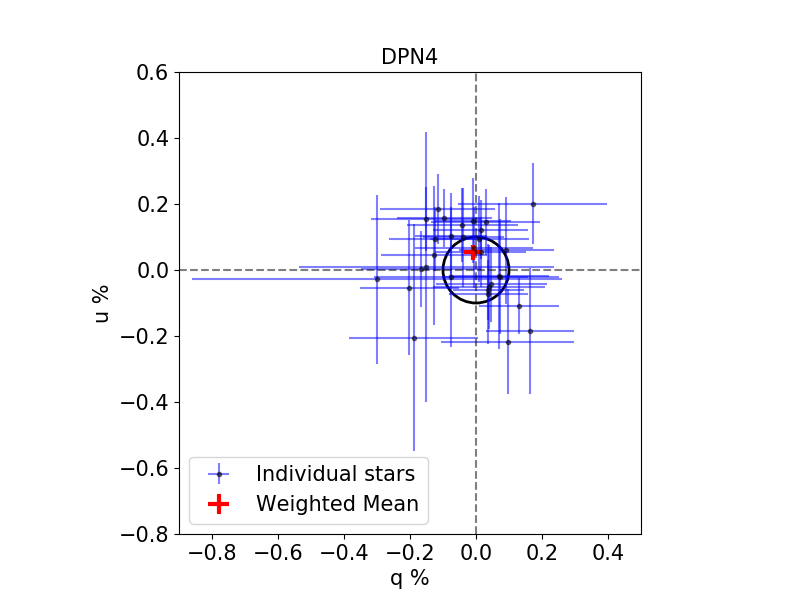}
        \label{fig:DPN4_qu}
    \end{subfigure}
    \hfill
    \begin{subfigure}[b]{0.49\textwidth}
        \centering
        \includegraphics[width=\textwidth]{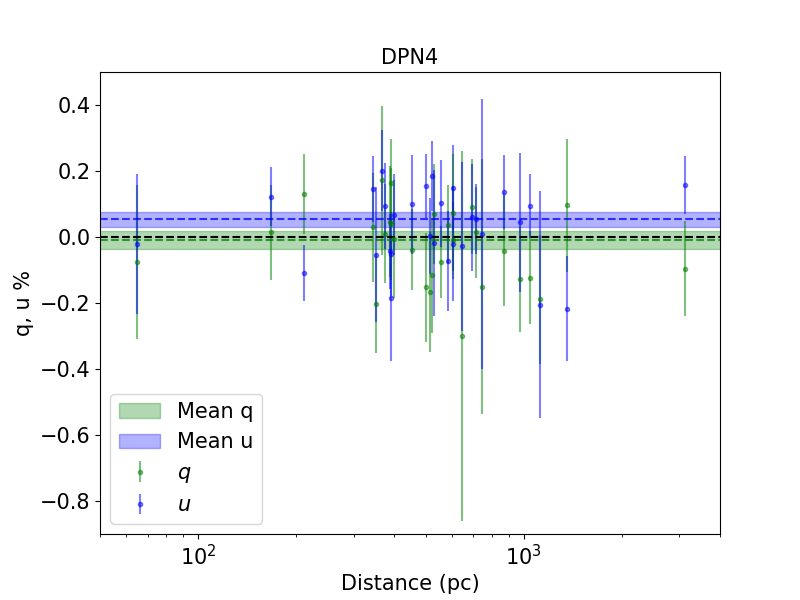}
        \label{fig:DPN4_dist}
    \end{subfigure}
    \caption{Same as Fig.~\ref{fig:DPN1} but for DPN4.}
    \label{fig:DPN4 }
\end{figure*}

\begin{figure*}[tbp]
    \centering
    \begin{subfigure}[b]{0.49\textwidth}
        \centering
        \includegraphics[width=\textwidth]{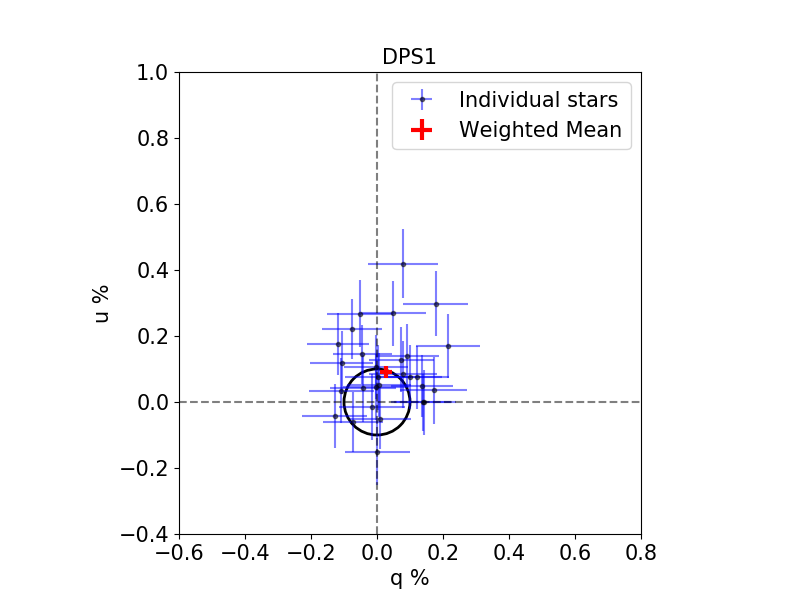}
        \label{fig:DPS1_qu}
    \end{subfigure}
    \hfill
    \begin{subfigure}[b]{0.49\textwidth}
        \centering
        \includegraphics[width=\textwidth]{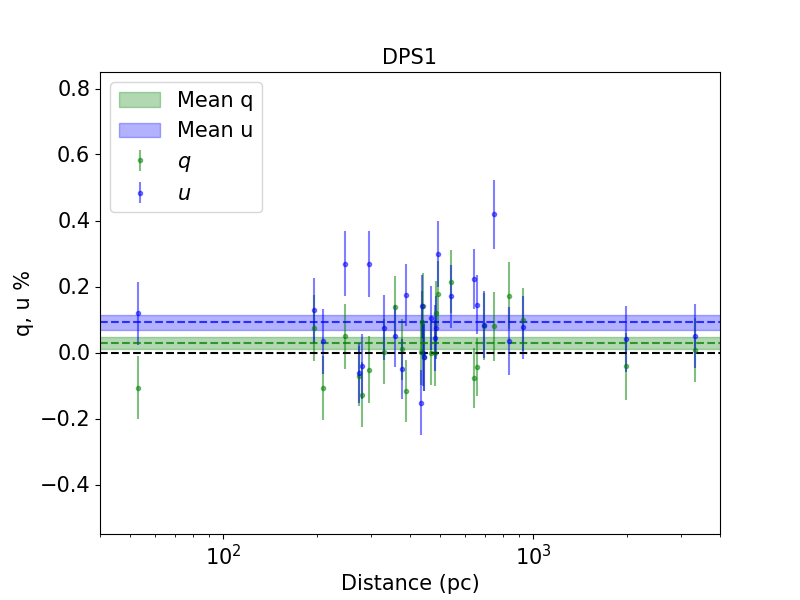}
        \label{fig:DPS1_dist}
    \end{subfigure}
    \caption{Same as Fig.~\ref{fig:DPN1} but for DPS1.}
    \label{fig:DPS1}
\end{figure*}

\begin{figure*}[tbp]
    \centering
    \begin{subfigure}[b]{0.49\textwidth}
        \centering
        \includegraphics[width=\textwidth]{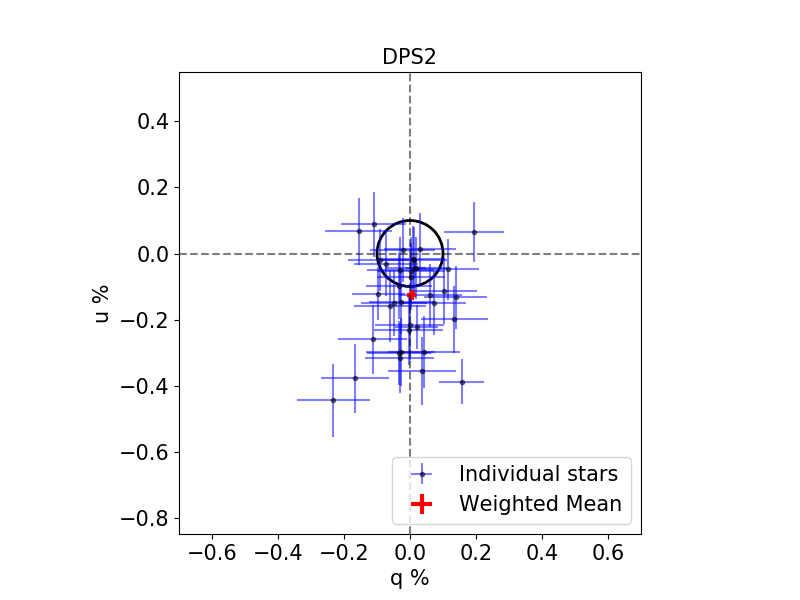}
        \label{fig:DPS2_qu}
    \end{subfigure}
    \hfill
    \begin{subfigure}[b]{0.49\textwidth}
        \centering
        \includegraphics[width=\textwidth]{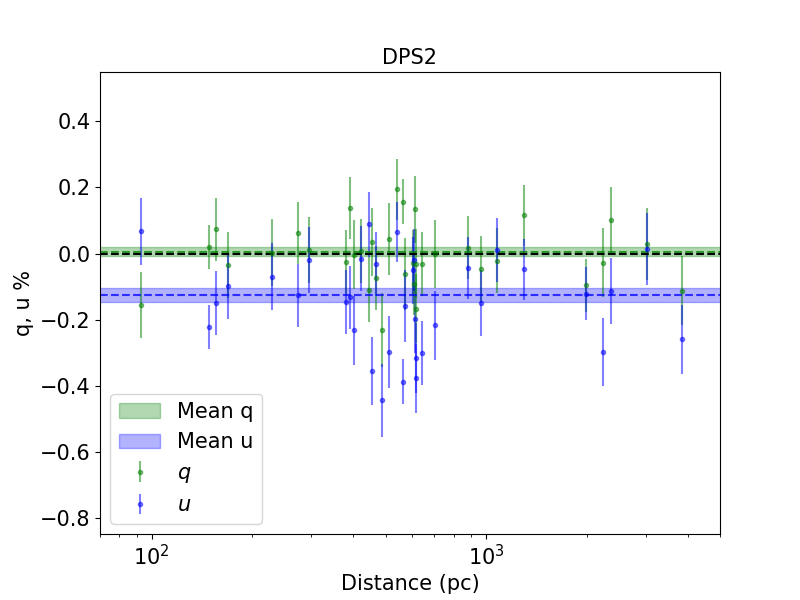}
        \label{fig:DPS2_dist}
    \end{subfigure}
    \caption{Same as Fig.~\ref{fig:DPN1} but for DPS2.}
    \label{fig:DPS2}
\end{figure*}

\begin{figure*}[tbp]
    \centering
    \begin{subfigure}[b]{0.49\textwidth}
        \centering
        \includegraphics[width=\textwidth]{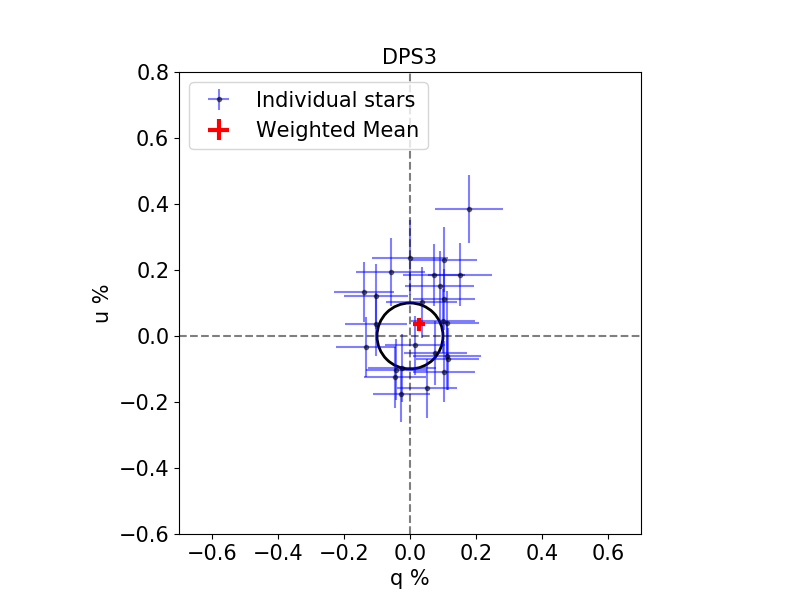}
        \label{fig:DPS3_qu}
    \end{subfigure}
    \hfill
    \begin{subfigure}[b]{0.49\textwidth}
        \centering
        \includegraphics[width=\textwidth]{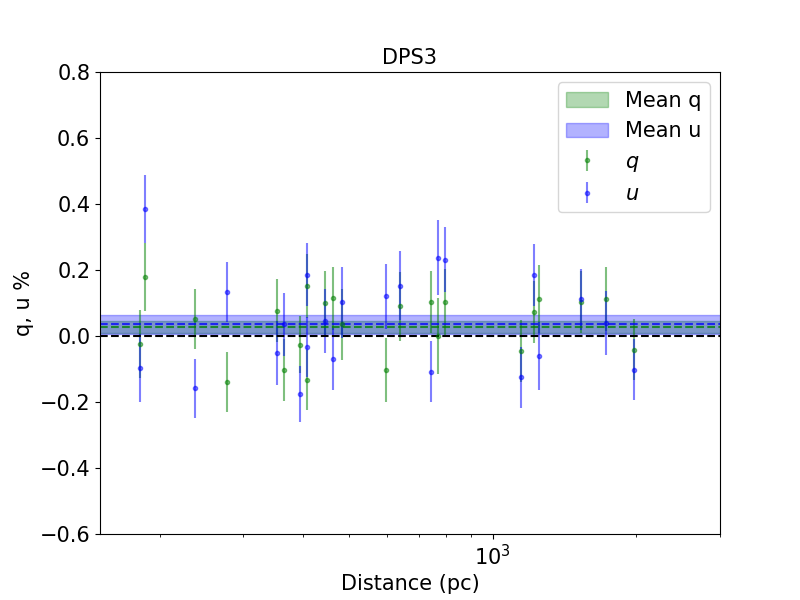}
        \label{fig:DPS3_dist}
    \end{subfigure}
    \caption{Same as Fig.~\ref{fig:DPN1} but for DPS3.}
    \label{fig:DPS3}
\end{figure*}

\begin{figure*}[tbp]
    \centering
    \begin{subfigure}[b]{0.49\textwidth}
        \centering
        \includegraphics[width=\textwidth]{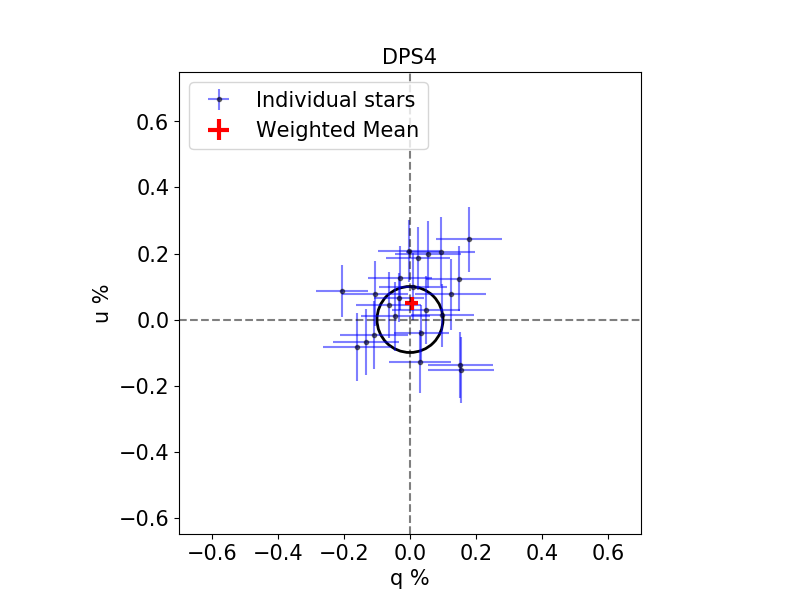}
        \label{fig:DPS4_qu}
    \end{subfigure}
    \hfill
    \begin{subfigure}[b]{0.49\textwidth}
        \centering
        \includegraphics[width=\textwidth]{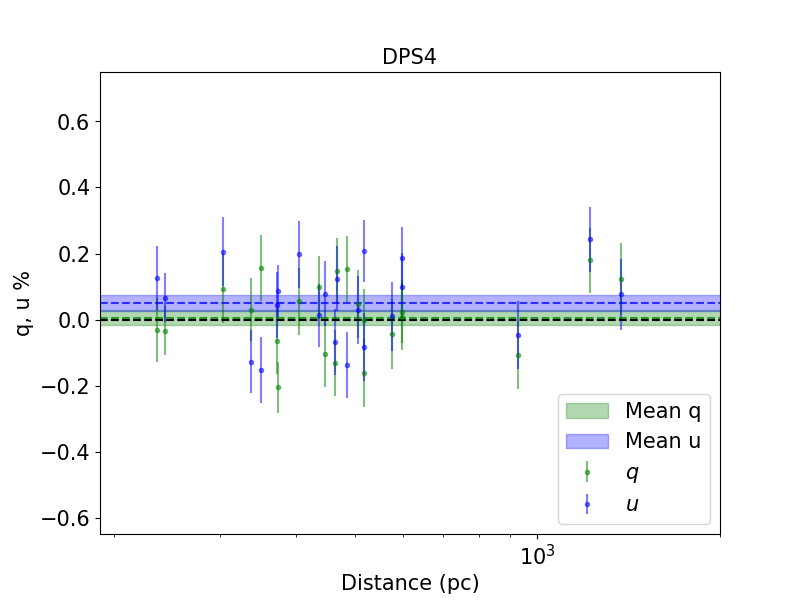}
        \label{fig:DPS4_dist}
    \end{subfigure}
    \caption{Same as Fig.~\ref{fig:DPN1} but for DPS4.}
    \label{fig:DPS4}
\end{figure*}

\begin{figure*}[tbp]
    \centering
    \begin{subfigure}[b]{0.49\textwidth}
        \centering
        \includegraphics[width=\textwidth]{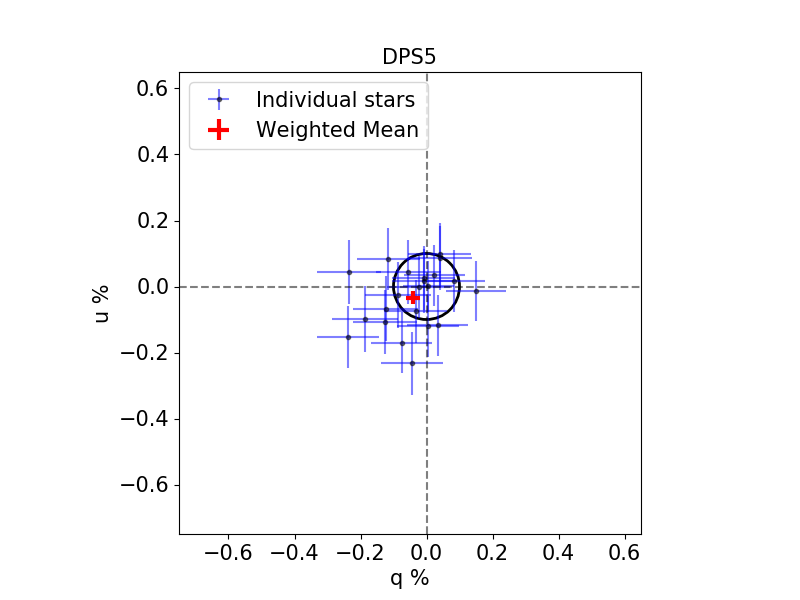}
        \label{fig:DPS5_qu}
    \end{subfigure}
    \hfill
    \begin{subfigure}[b]{0.49\textwidth}
        \centering
        \includegraphics[width=\textwidth]{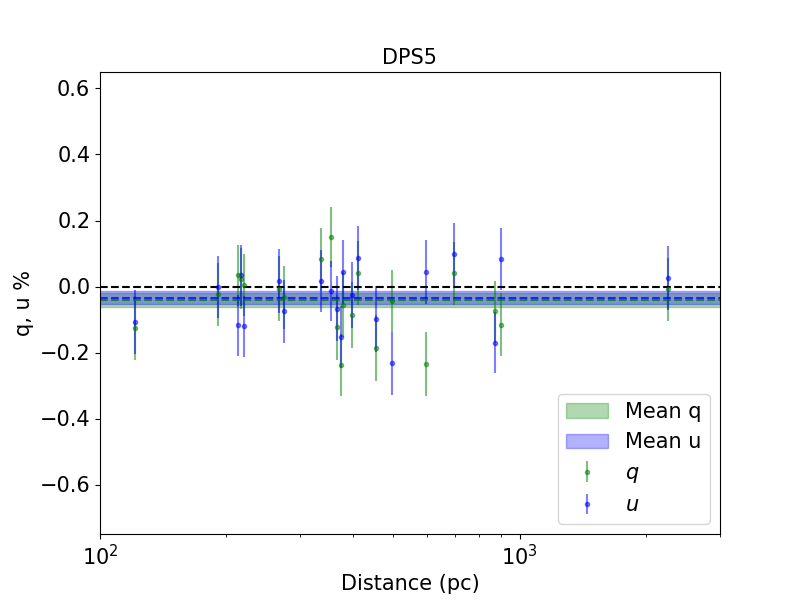}
        \label{fig:DPS5_dist}
    \end{subfigure}
    \caption{Same as Fig.~\ref{fig:DPN1} but for DPS5.}
    \label{fig:DPS5}
\end{figure*}

\begin{figure*}[tbp]
    \centering
    \begin{subfigure}[b]{0.49\textwidth}
        \centering
        \includegraphics[width=\textwidth]{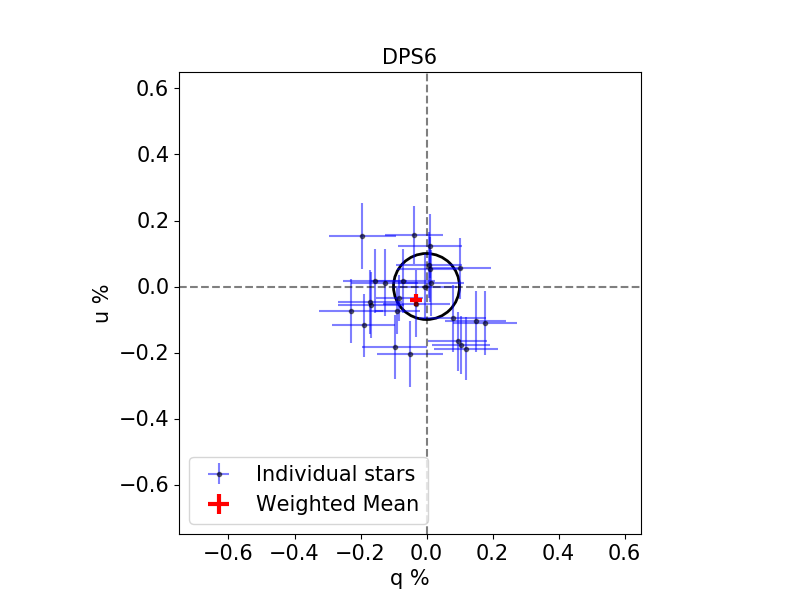}
        \label{fig:DPS6_qu}
    \end{subfigure}
    \hfill
    \begin{subfigure}[b]{0.49\textwidth}
        \centering
        \includegraphics[width=\textwidth]{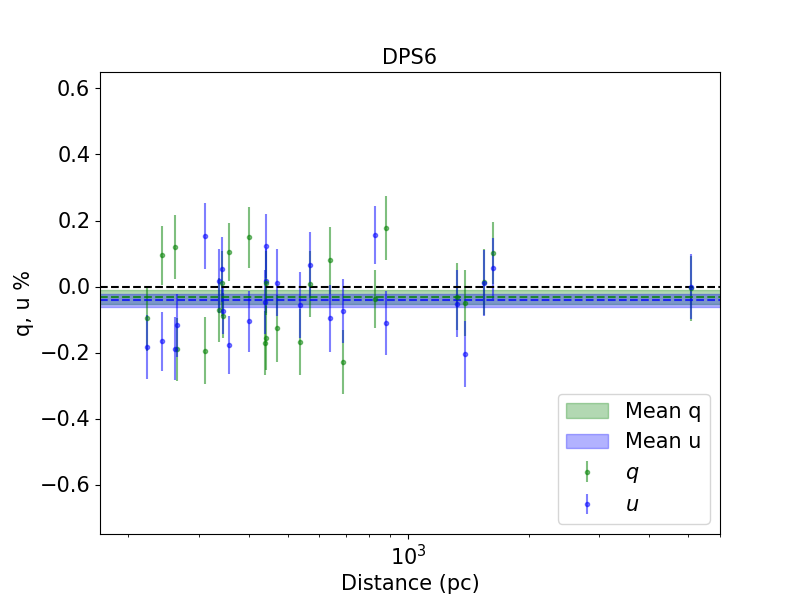}
        \label{fig:DPS6_dist}
    \end{subfigure}
    \caption{Same as Fig.~\ref{fig:DPN1} but for DPS6.}
    \label{fig:DPS6}
\end{figure*}

\begin{figure*}[tbp]
    \centering
    \begin{subfigure}[b]{0.49\textwidth}
        \centering
        \includegraphics[width=\textwidth]{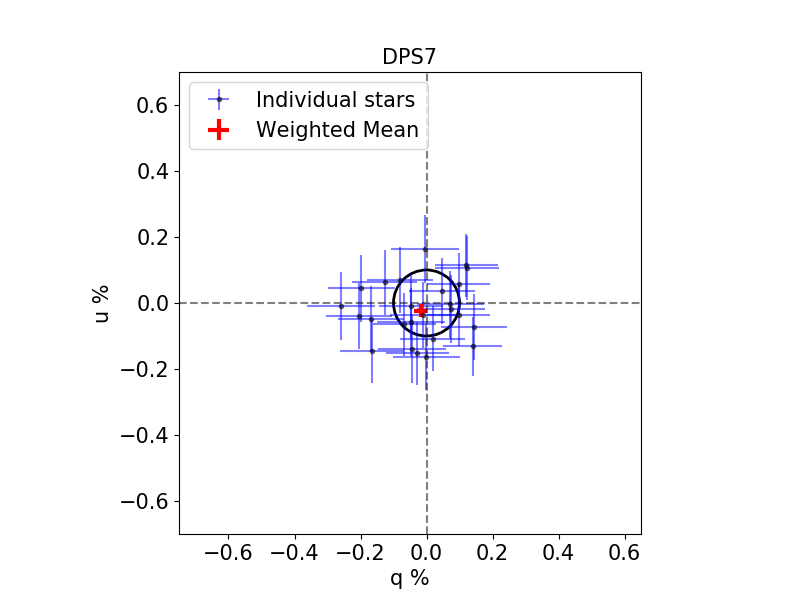}
        \label{fig:DPS7_qu}
    \end{subfigure}
    \hfill
    \begin{subfigure}[b]{0.49\textwidth}
        \centering
        \includegraphics[width=\textwidth]{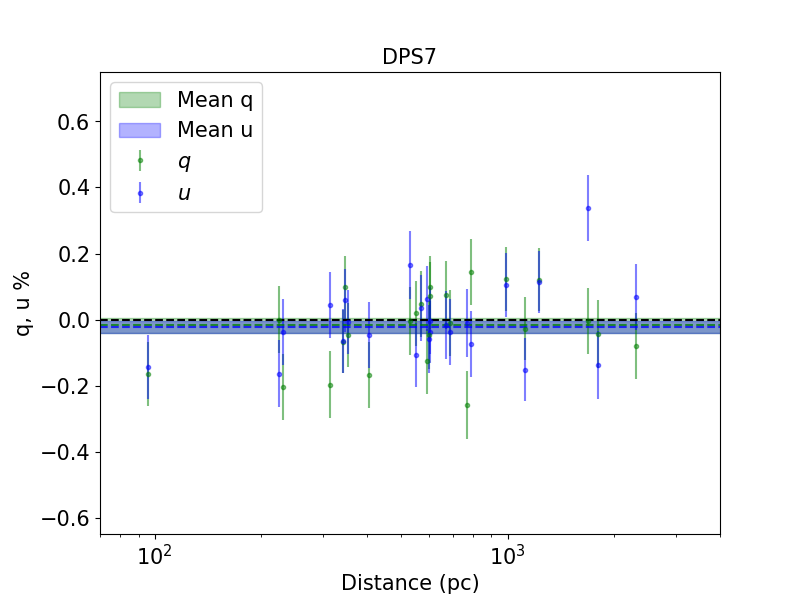}
        \label{fig:DPS7_dist}
    \end{subfigure}
    \caption{Same as Fig.~\ref{fig:DPN1} but for DPS7.}
    \label{fig:DPS7}
\end{figure*}

\begin{figure*}[tbp]
    \centering
    \begin{subfigure}[b]{0.49\textwidth}
        \centering
        \includegraphics[width=\textwidth]{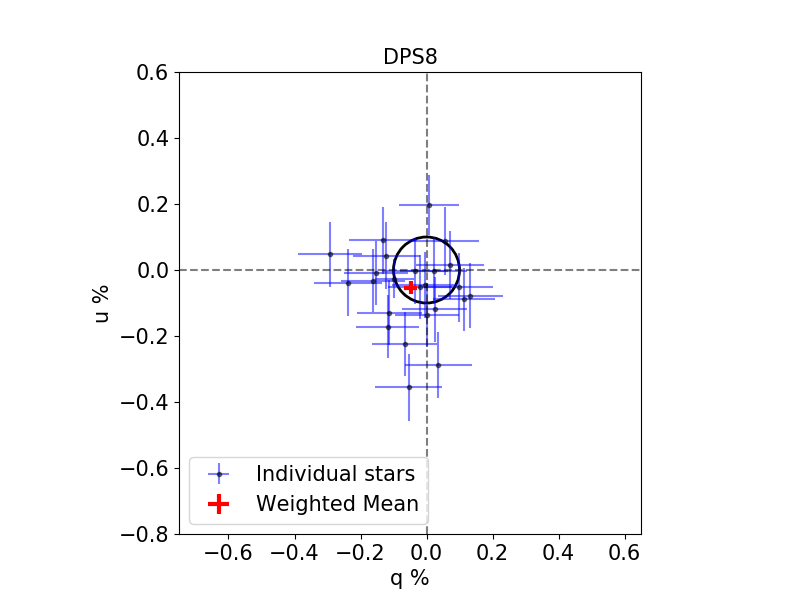}
        \label{fig:DPS8_qu}
    \end{subfigure}
    \hfill
    \begin{subfigure}[b]{0.49\textwidth}
        \centering
        \includegraphics[width=\textwidth]{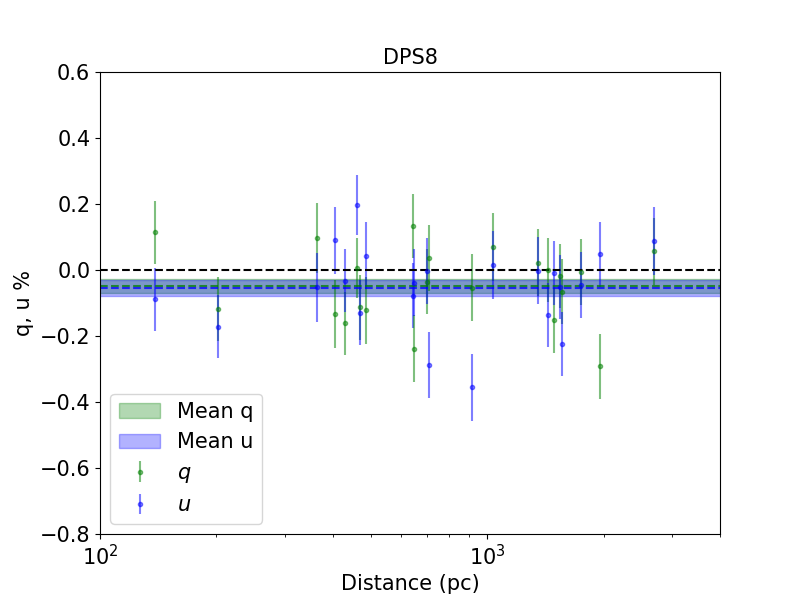}
        \label{fig:DPS8_dist}
    \end{subfigure}
    \caption{Same as Fig.~\ref{fig:DPN1} but for DPS8.}
    \label{fig:DPS8}
\end{figure*}

\FloatBarrier
\section{Finder charts of stars in the observed fields}\label{appendix:sky}

In this appendix, we present the sky maps of the dark patches, where we mark the observed stars. We also plot the dependence of the average polarization between different slices of RA and Dec of the same field to examine the presence of any spatial dependence of the polarization within the field.

\begin{figure}[htbp]
    \centering
    \includegraphics[width=0.49\textwidth]{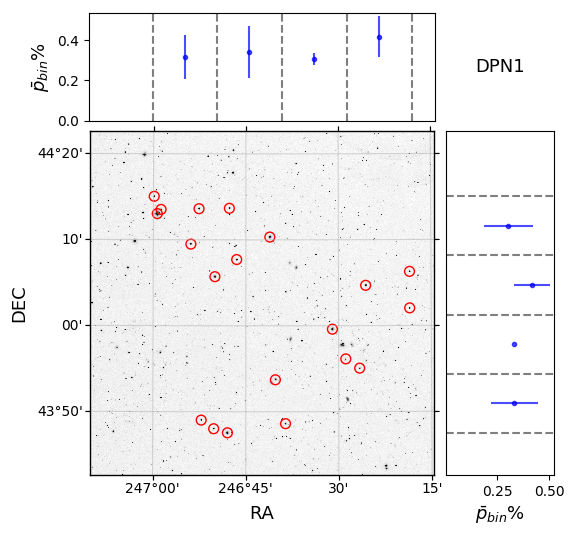}
    \caption{Observed stars in DPN1, marked with red circles. The RA and Dec range of the targets are divided in four equal bins. The blue points correspond to the average polarization of the stars within each bin. Errorbars represent the standard deviation of polarization measurements within each bin.}
    \label{fig:skyDPN1}
\end{figure}

\begin{figure}[htbp]
    \centering
    \includegraphics[width=0.49\textwidth]{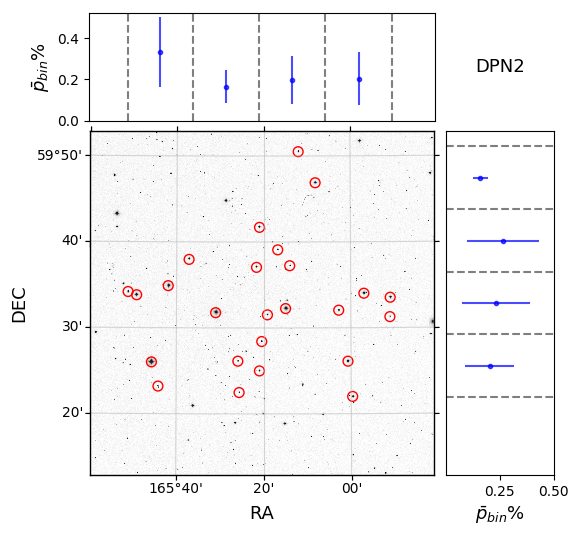}
    \caption{Same as Fig.~\ref{fig:skyDPN1} but for DPN2.}
    \label{fig:skyDPN2}
\end{figure}

\begin{figure}[htbp]
    \centering
    \includegraphics[width=0.49\textwidth]{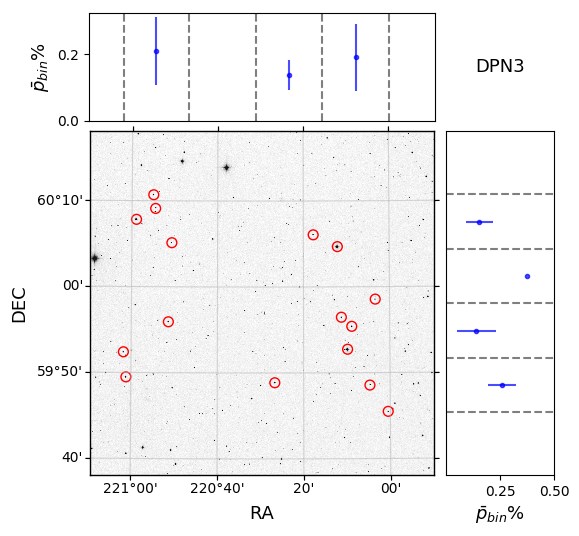}
    \caption{Same as Fig.~\ref{fig:skyDPN1} but for DPN3.}
    \label{fig:skyDPN3}
\end{figure}

\begin{figure}[htbp]
    \centering
    \includegraphics[width=0.49\textwidth]{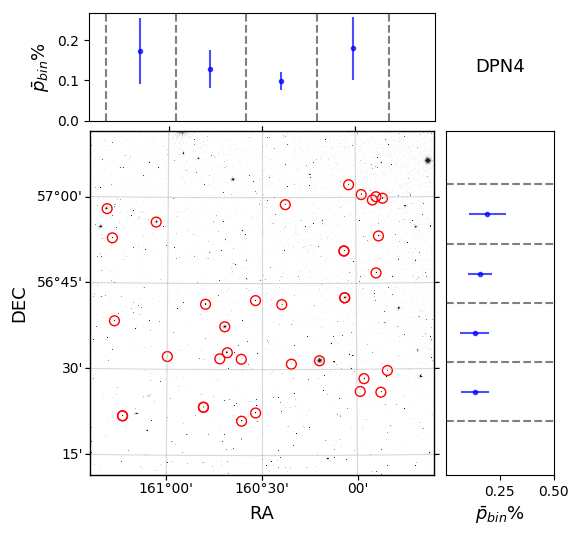}
    \caption{Same as Fig.~\ref{fig:skyDPN1} but for DPN4.}
    \label{fig:skyDPN4}
\end{figure}

\begin{figure}[htbp]
    \centering
    \includegraphics[width=0.49\textwidth]{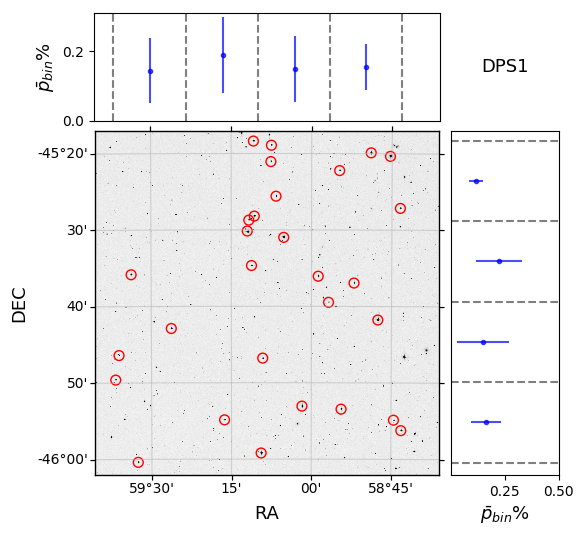}
    \caption{Same as Fig.~\ref{fig:skyDPN1} but for DPS1.}
    \label{fig:skyDPS1}
\end{figure}

\begin{figure}[htbp]
    \centering
    \includegraphics[width=0.49\textwidth]{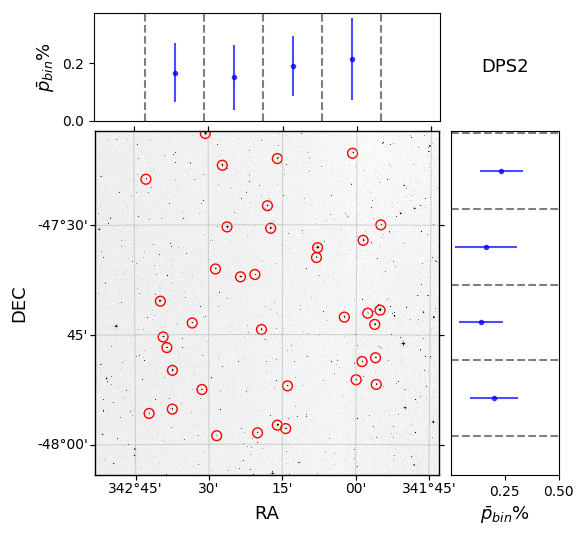}
    \caption{Same as Fig.~\ref{fig:skyDPN1} but for DPS2.}
    \label{fig:skyDPS2}
\end{figure}

\begin{figure}[htbp]
    \centering
    \includegraphics[width=0.49\textwidth]{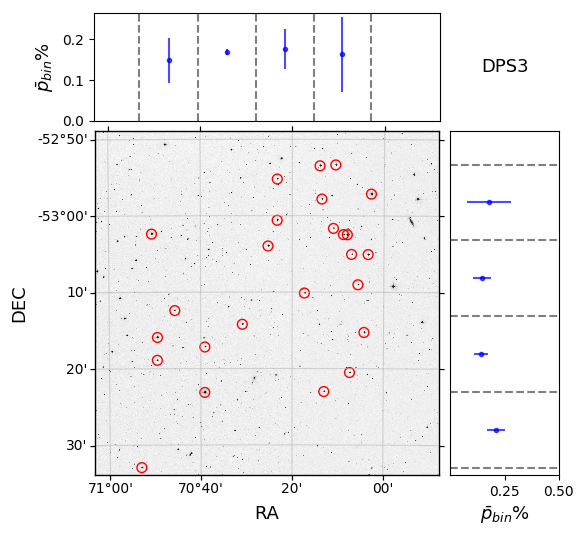}
    \caption{Same as Fig.~\ref{fig:skyDPN1} but for DPS3.}
    \label{fig:skyDPS3}
\end{figure}

\begin{figure}[htbp]
    \centering
    \includegraphics[width=0.49\textwidth]{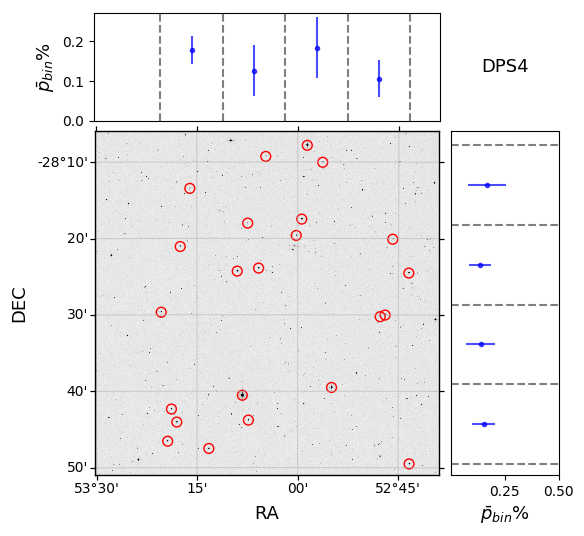}
    \caption{Same as Fig.~\ref{fig:skyDPN1} but for DPS4.}
    \label{fig:skyDPS4}
\end{figure}

\begin{figure}[htbp]
    \centering
    \includegraphics[width=0.49\textwidth]{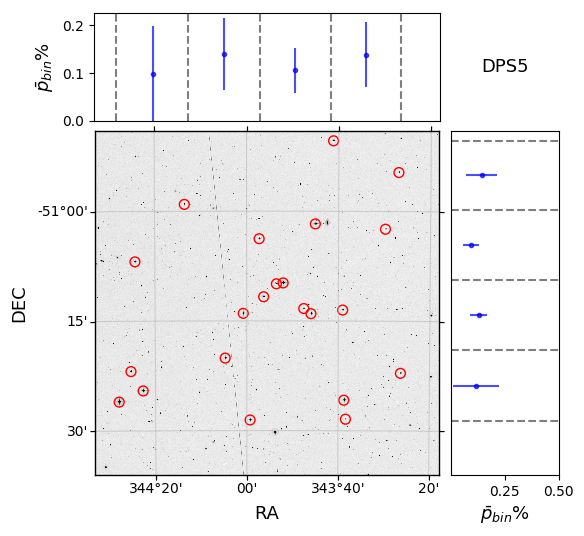}
    \caption{Same as Fig.~\ref{fig:skyDPN1} but for DPS5.}
    \label{fig:skyDPS5}
\end{figure}

\begin{figure}[htbp]
    \centering
    \includegraphics[width=0.49\textwidth]{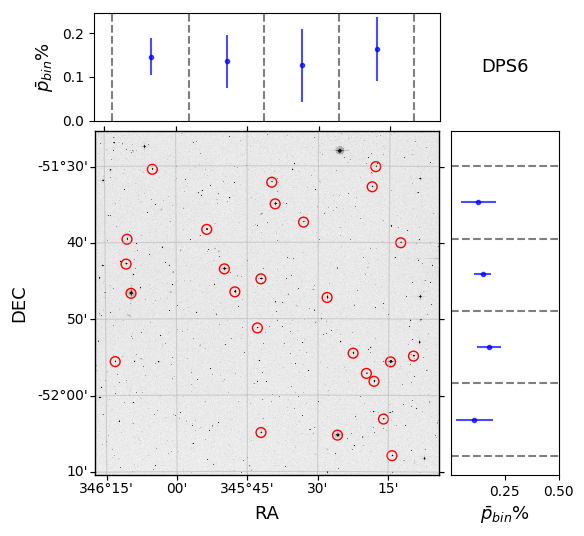}
    \caption{Same as Fig.~\ref{fig:skyDPN1} but for DPS6.}
    \label{fig:skyDPS6}
\end{figure}

\begin{figure}[htbp]
    \centering
    \includegraphics[width=0.49\textwidth]{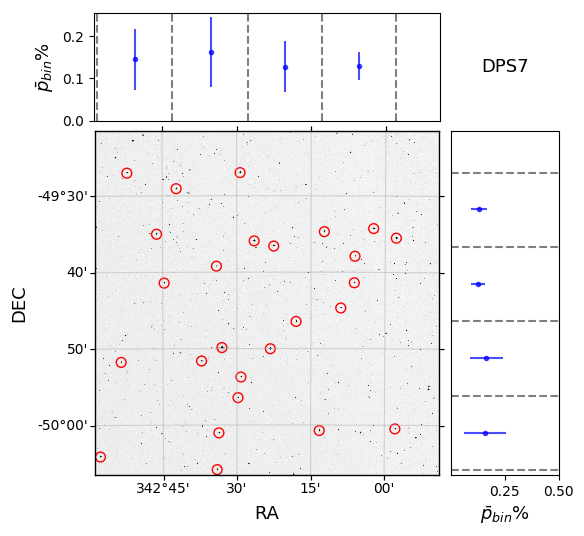}
    \caption{Same as Fig.~\ref{fig:skyDPN1} but for DPS7.}
    \label{fig:skyDPS7}
\end{figure}

\begin{figure}[htbp]
    \centering
    \includegraphics[width=0.49\textwidth]{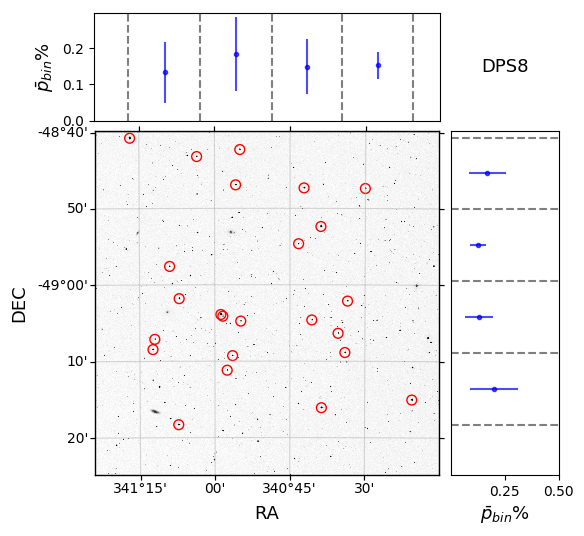}
    \caption{Same as Fig.~\ref{fig:skyDPN1} but for DPS8.}
    \label{fig:skyDPS8}
\end{figure}

\end{document}